\documentclass{emulateapj}
\newcommand{\KMS}{\mbox{km s}^{-1}}    

\def\olap#1#2#3#4{\langle #1, #2 | #3, #4 \rangle}
\def\Vkick{$V = 475\,\KMS\,a\,$}

\def\ltsima{$\; \buildrel < \over \sim \;$}
\def\ltsim{\lower.5ex\hbox{\ltsima}}
\def\gtsima{$\; \buildrel > \over \sim \;$}
\def\gtsim{\lower.5ex\hbox{\gtsima}}

\def\MPR#1{{\it Moving Puncture Recipe}#1 (MPR#1)\gdef\MPR{MPR}}
\def\ahz#1{apparent horizon#1 (AH#1)\gdef\ahz{AH}}
\def\CLA#1{close-limit approximation#1 (CLA#1)\gdef\CLA{CLA}}
\def\pnw#1{post-Newtonian#1 (PN#1)\gdef\pnw{PN}}
\def\qnm#1{quasi-normal mode#1 (QNM#1)\gdef\qnm{QNM}}
\def\isco#1{innermost stable circular orbit#1 (ISCO#1)\gdef\isco{ISCO}}
\def\eos#1{equation of state#1 (EOS#1)\gdef\eos{EOS}}
\def\ns#1{neutron star#1 (NS#1)\gdef\ns{NS}}
\def\bbh#1{binary black holes#1 (BBH#1)\gdef\bbh{BBH}}
\def\bhns#1{black hole -- neutron star#1 (BHNS#1)\gdef\bhns{BHNS}}
\def\nsns#1{neutron star -- neutron star#1 (NSNS#1)\gdef\nsns{NSNS}}
\def\emri#1{extreme mass-ratio inspiral#1 (EMRI#1)\gdef\emri{EMRI}}
\def\emrb#1{extreme mass-ratio binaries#1 (EMRB#1)\gdef\emrb{EMRB}} 
\def\grb#1{gamma-ray burst#1 (GRB#1)\gdef\grb{GRB}}
\def\imbh#1{intermediate mass black hole#1 (IMBH#1)\gdef\imbh{IMBH}}
\def\smbh#1{supermassive black hole#1 (SMBH#1)\gdef\smbh{SMBH}}
\def\bh#1{black hole#1 (BH#1)\gdef\bh{BH}}
\def\ulx#1{ultra-luminous x-ray source#1 (ULX#1)\gdef\ulx{ULX}}
\def\lmxbs{low-mass x-ray Binaries (LMXBs)\gdef\lmxbs{LMXBs}\gdef\lmxb{LMXB}} 
\def\lmxb{low-mass x-ray Binary (LMXB)\gdef\lmxbs{LMXBs}\gdef\lmxb{LMXB}} 

\begin{document}
\title{Gravitational recoil from spinning binary black hole mergers}

\author{Frank Herrmann, Ian Hinder, Deirdre Shoemaker\altaffilmark{1}, Pablo
Laguna\altaffilmark{2}}

\affil{Center for Gravitational Wave Physics, \\
The Pennsylvania State University,
University Park, PA 16802, USA}

\author{Richard A. Matzner}

\affil{Center for Relativity and Department of Physics\\
The University of Texas at Austin, Austin, Texas 78712, USA}

\altaffiltext{1}{IGPG, Department Physics,
The Pennsylvania State University, University Park, PA 16802, USA}

\altaffiltext{2}{IGPG, Departments of Astronomy \& Astrophysics and Physics,
The Pennsylvania State University, University Park, PA 16802, USA}

\begin{abstract}
  The inspiral and merger of binary black holes
  will likely involve
  black holes with both unequal masses and arbitrary spins. The
  gravitational radiation emitted by these binaries will carry angular
  as well as linear momentum.  A net flux of emitted linear momentum
  implies that the black hole produced by the merger will experience a
  recoil or kick.  Previous studies have focused on the recoil
  velocity from unequal mass, non-spinning binaries.  We present results from
  simulations of equal mass but spinning black hole binaries and show
  how a significant gravitational recoil can also be obtained in these
  situations.  We consider the case of black holes with opposite spins
  of magnitude $a$ aligned/anti-aligned with the orbital angular momentum,
  with $a$ the dimensionless spin parameters of the individual holes.
  For the initial setups under consideration, we find a recoil velocity of \Vkick.
  Supermassive black hole mergers producing kicks of this
  magnitude could result in the ejection from the cores of
  dwarf galaxies of the final hole produced by the
  collision.
\end{abstract}

\keywords{
          black hole physics ---
          galaxies: nuclei ---
          gravitation ---
          gravitational waves ---
          relativity
}

\section{Introduction}\label{sec:intro}

There is ample observational evidence that \smbh{s} are common at the
centers of galaxies~\citep{Richstone:1998ea,Magorrian:1997hw}, with
masses in the range $10^5 - 10^9 M^{}_{\odot}$. These \smbh{s} are involved in
exciting astrophysical phenomena.  For instance, there is a remarkable, not
completely understood, correlation between the velocity
dispersion of the bulge of the host galaxy and the mass of the
\smbh{}~\citep{Ferrarese:2000fm}.  There is also indication 
of a correlation of the mass of the 
\smbh{} with the mass
of the host dark matter halo~\citep{Ferrarese:2002la}.  An interesting
aspect of \smbh{ } growth arises as a consequence of hierarchical cold
dark matter cosmologies, in which large-scale structures are formed by
mergers.  \smbh{s} would then grow both by gas accretion and by
coalescence with other \smbh{s} (brought together when their host
galaxies collide~\citep{Volonteri:2002am,Begelman:1980vb}).  The work
in this paper focuses on one aspect of the merger of \smbh{s}, the
kick in the final \smbh{}.

The late inspiral and merger of \smbh{s} produces extremely energetic
gravitational radiation, which will be observable by the planned
space-based gravitational wave antenna
LISA~\citep{Danzmann:2003ad,Prince:2003aa}.  Gravitational radiation
produced during the inspiral and merger of \bh{s} not only carries
energy with it, but, except in special-symmetry cases,
can also transport net linear and angular momentum.
For instance, in
the merger of unequal mass \smbh{s}, a net flux of linear momentum will
be emitted by the system~\citep{Peres:1962ap,Bekenstein:1973jd}.  As a
consequence, the final \bh{} will experience a gravitational recoil or
kick.  There are observations that hint at such scenarios, in which a
\smbh{} has been ejected in an ongoing galaxy
merger~\citep{Haehnelt:2006hd}. (An alternative explanation could be
that the ejection is due to gravitational slingshot of three or more
\smbh{s} in the merger.) It is then very important to get good
estimates of recoil velocities in \bh{} mergers. These estimates have
a profound effect on the understanding of the demographics of \smbh{s}
at the cores of galaxies, their growth~\citep{Haiman:2004ve} and their
merger rates~\citep{Micic:2006ta}.  Knowledge of the conditions under
which kicks are produced could also help explain the absence of
massive \bh{s} in dwarf galaxies and stellar
clusters~\citep{Madau:2004mq,Merritt:2004xa}, and could determine the
population of \bh{s} in the interstellar and intergalactic medium.

Gravitational recoil estimates of unequal mass binaries have been
addressed using both analytic and full numerical relativity
approaches.  The first quasi-Newtonian analytic studies
~\citep{Fitchett:1983fc,Fitchett:1984fd} produced kick velocities as
large as $\sim 1500\, \KMS$.  \citet{Wiseman:1992dv} and more recently
\citet{Blanchet:2005rj} and \citet{Damour:2006tr} improved these
estimates by including \pnw{} effects. The maximum kick in these
studies was found to be in the range of $\sim 74 - 250\, \KMS$, and it
occurred for $\eta \sim 0.2$, where $\eta \equiv M_1M_2/(M_1+M_2)^2$
is the symmetrized mass ratio parameter.  (This corresponds to 
a mass ratio $q \equiv M_1/M_2 \sim 0.38$.) These analytic \pnw{}
studies also showed that the final value of the kick is mostly
accumulated during the merger or plunge phase of the binary.  Since
the plunge phase is beyond the limit of applicability of \pnw{}
approximations, the results can only be taken as ``best-bet
estimates''~\citep{Damour:2006tr}.

There are two semi-analytic studies that in principle had a better
handle on the plunge phase.  \citet{Campanelli:2004zw} obtained kick
velocities of $\sim 300\, \KMS$ using the {\em Lazarus} approach, a
framework~\citep{Baker:2001sf} that combines full numerical relativity
and \CLA{} perturbation theory~\citep{1994PhRvL..72.3297P}.  More
recently, \citet{2006astro.ph.11110S} and \citet{2006PhRvD..74l4010S}
combined \pnw{} estimates during the inspiral with kick estimates
using the \CLA{.}  The maximum recoil obtained in this work was $\sim 167
(1 + e) \KMS$, with $e$ the eccentricity of the binary.  Finally, full
numerical relativity studies have also been carried out by \citet{Herrmann:2006ks},
\citet{Baker:2006vn} and
\citet{Gonzalez:2006md}. Only full numerical relativity approaches
provide accurate estimates of kicks since they correctly handle the
non-linear behavior of the plunge.  The most comprehensive study so
far is that by \citet{Gonzalez:2006md}, in which a maximum kick
velocity of $\sim 175\, \KMS$ was obtained also for $\eta \sim 0.2$ 
$(q \sim 0.38$), consistent with \pnw{} studies.  What
is interesting is that the findings of \citet{2006astro.ph.11110S}
based on the \CLA{} are remarkably close to the full
numerical relativity results by \citet{Gonzalez:2006md}, 
supporting the view that the kick is mostly due to the linear momentum
emitted during the plunge, where the \CLA{} has been demonstrated to
provide a good approximation \citep{1995PhRvD..52.4462A}.  
To our knowledge, the only kick study involving spinning \bh{s} is 
that by~\citet{Favata:2004wz}. 
They considered
the case of an extreme-mass-ratio system with a spinning
\smbh{}. Using \bh{} perturbation theory they estimated kick
velocities of $\sim 100 - 200\, \KMS$. 
Head-on collisions of spinning \bh{s} have also been recently
considered~\citep{Dale07}.

To help us understand {\it our} computational results, we present next a
rough order-of-magnitude estimate of the kicks one should expect.
Note first that there must be some asymmetry between the \bh{s} in
order for there to be asymmetric radiation which can lead to
kicks. Thus, in the non-spinning case, unequal masses are required;
here we consider binaries of equal masses, but different spin
(magnitude or direction).  The kick is expected to increase as
the relevant spin increases, but especially symmetric cases will still
show zero kick (e.g.~when the \bh{s} have their spins aligned parallel to the 
orbital angular momentum).  The
order-of-magnitude estimate can be obtained from the radiative
linear momentum loss formula
\citep{RevModPhys.52.299,PhysRevD.52.821}.  Excluding non-spin
terms, this formula reads
\begin{eqnarray}
\label{kick}
\frac{dP^i}{dt} &=& \frac{16}{45} \epsilon^{ijk} I^{(3)}_{jl}
H^{(3)}_{kl}
+ \frac{4}{63} H^{(4)}_{ijk} H^{(3)}_{jk}\nonumber \\
&+& \frac{1}{126} \epsilon^{ijk} I^{(4)}_{jlm} H^{(4)}_{klm}\,.
\end{eqnarray}
Here $I_{ij}$ and $I_{ijk}$ are respectively the mass quadrupole and
octupole.  Similarly, $H_{ij}$ and $H_{ijk}$ are the spin quadrupole
and octupole, respectively.  In (\ref{kick}), a super-index $\,^{(n)}$
denotes an $n$th-time derivative.  Clearly equation (\ref{kick}) predicts
a periodic force for exactly circular orbits.  As the \bh{s} spiral
together the strength of the periodic kick increases, so we estimate
the kick from the last half orbit before merger.

Consider a binary system consisting of \bh{s} in circular orbit 
with equal masses ($M_1 = M_2 = M/2$). In the absence of spin 
this would produce no kick, but here we set data with each \bh{} 
having spin perpendicular to the orbit, the spins oppositely directed, 
each with dimensionless Kerr spin parameter $a$ ($0 \le a\le 1$).
This is the configuration we use below for our
computational evaluation of the kick.
The calculation of the mass quadrupole is familiar,
and for circular orbits in the $xy$ plane with orbital angular
velocity $\omega$ and coordinate separation $d$ gives nonzero values:
\begin{eqnarray}
I^{(3)}_{xx} &=&   2M\,d^2\,\omega^3 \sin{(2\omega t)}\nonumber\\
I^{(3)}_{xy} &=& - 2M\,d^2\,\omega^3 \cos{(2\omega t)}\nonumber\\
I^{(3)}_{yy} &=& - 2M\,d^2\,\omega^3 \sin{(2\omega t)}\,.
\label{massQP}
\end{eqnarray}
The spin quadrupole can be most easily calculated by imagining a spin dipole
(charges $\pm M/2$, separation $a\,M/2$) and conceptually taking the limit
at the end.
The result is 
\begin{eqnarray}
H^{(3)}_{xz} &=& \frac{1} {4}\, M^2 d\,a\,\omega^3 \sin{(\omega t)}\nonumber\\
H^{(3)}_{yz} &=& -\frac{1} {4}\, M^2 d\,a\,\omega^3 \cos{(\omega t)}\,.
\label{spinQP}
\end{eqnarray}
Inserting (\ref{massQP}) and (\ref{spinQP}) into the first term in
equation~(\ref{kick}) gives
\begin{eqnarray}
\frac{dP^x}{dt} &=& \frac{8} {45} M^3 d^3a\, \omega^6 \sin{( \omega t)}
\nonumber\\
\frac{dP^y}{dt} &=& -\frac{8} {45} M^3 d^3a\, \omega^6 \cos{( \omega t)} \,.
\label{Pxy}
\end{eqnarray}
Notice that the force is in the plane of the orbit and 
rotates with the orbit.  The average over half a
cycle is $2/ \pi$, so equation (\ref{Pxy}) is a good estimate for any half
cycle as the orbit spirals in. The total force can then be
approximated as
\begin{equation}
\frac{dP}{dt} = \frac{16} {45\,\pi} M^3 d^3a\, \omega^6 \,.     
\end{equation}
Compare this to the total luminosity:
\begin{eqnarray}
\frac{dE}{dt} &=& \frac{2}{5} M^2 d^4 \omega^6 \,.
\label{Py}
\end{eqnarray}
Thus the asymmetry in radiation that contributes to the kick is
\begin{eqnarray}
\frac{dP}{dE} = \frac{dP}{dt} \big{/} \frac{dE}{dt}&=& \frac{4\,a}{9\,\pi}\frac{M}{d}\,,
\end{eqnarray}
which is $\sim 0.02$ for dimensionless spin parameter $a\sim 1/2 $ and $d \sim 6 M$.
(The latter is an estimate of the separation near the ``last orbit''.)
If $\Delta E$ is the total energy radiated by the binary, an estimate of the
(half orbit) kick is
\begin{eqnarray}
V &\sim& c \left(\frac{dP}{dE}\right)\left(\frac{\Delta E}{M}\right)\nonumber\\
&\sim& 300 \KMS \left(\frac{dP/dE}{0.02}\right)\left(\frac{\Delta E/M}{0.05}\right) \,.
\end{eqnarray}

For another estimate, we note that ~\citet{Favata:2004wz} specialized the \pnw{} equation~(3.31) in
~\citet{Kidder:1995zr} to the case of circular orbit with spins parallel and anti-parallel
to the orbital angular momentum. The resultant kick velocity is given by
\begin{equation}
V = V_q + 883\KMS \left(\frac{f_{SO}(q,a1,a2)}{f_{SO,\mathrm{max}}}\right)
\left(\frac{2\,M}{r_{\mathrm{term}}}\right).
\label{eq:kidder}
\end{equation}
Above $V_q$ is the contribution to the kick that depends only on the
mass ratio $q$; this contribution vanishes
for equal mass binaries ($q=1$). The radius $r_{\mathrm{term}}$ is the separation
at which gravitational radiation terminates. The scaling function in equation~(\ref{eq:kidder})
is given by $f_{SO}(q,a_1,a_2) = q^2(a_2-q\,a_1)/(1+q)^5$ with
$f_{SO,\mathrm{max}} = f_{SO}(1,\pm1,\mp1) = 1/16$. Therefore, for the cases 
we have investigated, equation (\ref{eq:kidder}) reduces to
$V = 883\,\KMS a\, (2\,M/r_{\mathrm{term}})$, comparable to our estimate above, for reasonable choices of $r_{\mathrm{term}}$.

There is another effect similar to the pulsar kick mechanism described
by~\citet{1975ApJ...201..447H}. It involves explicit retardation
effects (so is not captured in the multipole expression
of equation~(\ref{kick})), and gives estimates of similarly sized kicks.
We shall see that full numerical relativity simulations give
comparable kicks to this estimate.

The paper is organized as follows: In Sec.~\ref{sec:movingpunc}, we
present the computational methodology and details of how the initial
data were constructed. Sec.~\ref{sec:recoil} gives details of
the method to estimate kicks. Code tests and a convergence analysis are
given in Sec.~\ref{sec:converge}. The gravitational recoil estimates
are presented in Sec.~\ref{sec:results}, with 
conclusions given in Sec.~\ref{sec:discussion}.

\section{Computational Methodology and Initial Data} 
\label{sec:movingpunc}

The numerical simulations of \bbh{} in our work were obtained
following the \MPR{}. The essence of this recipe is: (A) a particular
formulation of the Einstein field equations and (B) a set of
coordinate or gauge conditions for updating field variables during
evolution as well as for handling the \bh{} singularities.  The form
of the evolutions required by the \MPR{} is the so-called BSSN 3+1
formulation of Einstein's equations~\citep{Nakamura87, Shibata95,
  Baumgarte99}.  A derivation of the BSSN equations and a few examples
of their applications can be found in the review
by~\citet{Baumgarte:2002jm}.

In addition to the form of the evolution equations, the success of the
\MPR{} is due to the coordinate or gauge
conditions~\citep{Alcubierre02a,Baker:2005vv,Campanelli:2005dd}.  The \MPR{} gauge
conditions are equations that determine the lapse function $\alpha$
and the shift vector $\beta^i$.  The lapse is a ``local'' measure of
proper time, and the shift vector encapsulates the freedom of labeling
events at a given time~\citep{Baumgarte:2002jm}.  The explicit form of
the evolution equations for the lapse and shift 
in the \MPR{} are $\partial_0\alpha = - 2\alpha K$ 
and for the shift $\partial_0 \beta^i= 3/4 B^i$ and 
$\partial_0 B^i = \partial_0 \widetilde{\Gamma}^i -\xi B^i$, where 
$\partial_0 = \partial_t - \beta^j \partial_j$. $K$
is the trace of the extrinsic curvature, $\widetilde\Gamma^i$ is the
trace of the conformal connection and $\xi=2$ is a free dissipative
parameter.  The importance of these gauge conditions is twofold:
first, they avoid the need of excising the \bh{} singularity from the
computational domain since they effectively halt the evolution
(i.e. the lapse function $\alpha$ vanishes) near the \bh{}
singularity~\citep{Hannam:2006vv}. Second, they allow for movement of
the \bh{} or {\it puncture} throughout the computational domain while
freezing the evolution inside the \bh{} horizon.

The code used for this work was produced by the \texttt{Kranc} code
generation package~\citep{Husa:2004ip}, the \texttt{Cactus}
infrastructure~\citep{Cactusweb} for parallelization and
\texttt{Carpet}~\citep{Schnetter-etal-03b} for mesh refinement.  The
code is based on fourth order accurate finite differencing of
spatial operators and uses 4th order Runge-Kutta for time integration with a 
Courant factor of 0.5.

The initial data use punctures~\citep{Brandt97b} to represent \bh{s}.
In Einstein's theory, initial data are not completely freely
specifiable; they must satisfy the Hamiltonian and momentum
constraints.  We use the spectral code
developed by~\citet{Ansorg:2004ds} to solve these constraints.  The
initial free-data (e.g.~angular momentum, spins, masses, separations)
are chosen according to the effective potential
method~\citep{Cook94,Baumgarte00a}.  This method yields \bbh{} initial
data sets representing \bbh{s} in quasi-circular orbit. In general
terms, the effective potential method consists of minimizing the
``binding energy'' of the binary to determine the \bbh{} parameters.

\begin{deluxetable}{l c c c c c c }
\tablewidth{0pt}
\tablecaption{\label{tab:initial_data}Initial Data Parameters}
\tablehead{
\colhead{Model} & \colhead{$x/M$} & \colhead{$P/M$} & \colhead{$S/M^2$}
& \colhead{$m_1/M$} & \colhead{$m_2/M$} &\colhead{$E/M$} }
\startdata
S0.05  & 2.95  & 0.13983 & 0.05 &  0.4683 & 0.4685 & 0.98445\\
S0.10  & 2.98  & 0.13842 & 0.10 &  0.4436 & 0.4438 & 0.98455\\
S0.15  & 3.05  & 0.13547 & 0.15 &  0.3951 & 0.3953 & 0.98473\\
S0.20  & 3.15  & 0.13095 & 0.20 &  0.2968 & 0.2970 & 0.98499\\
\enddata
\end{deluxetable}

Table~\ref{tab:initial_data} contains the \bbh{} parameters of our
simulations.  The \bh{s} are located at positions $(\pm x/M,0,0)$,
have linear momentum $(\pm P/M,0,0)$, spin $(0,0,\pm S/M^2)$ and bare
puncture masses $m_{1,2}/M$, with $M=M_1+M_2$ the total mass of the
binary. Notice that the bare puncture masses are slightly different.
The reason for this difference is because of the spin contribution to
the mass of each hole (measured from the area of their apparent horizons);
in order to keep the individual masses of the \bh{s}, $M_1$ and $M_2$,
equal, (slight) adjustments to the bare masses are necessary.  
The configurations are such that the total angular momentum
is for all cases $J/(\mu M) = 3.3$ with $\mu = M_1M_2/(M_1+M_2)$.
It is important to notice that $S$ is not the Kerr spin parameter $0\le a_{Kerr}\le
m_{BH}$ typically associated with rotating \bh{s}. The dimensionless spin parameter for
each \bh{} is given by $a_{1,2}=S/M_{1,2}^2$ with $M_{1,2}=M/2$. 
The cases considered here, $S/M^2 = \lbrace 0.05, 0.10, 0.15, 0.20 \rbrace$,
correspond to $a = \lbrace 0.2, 0.4, 0.6, 0.8 \rbrace$, respectively. 
For reference, the total ADM mass $E/M$ in the initial data 
is also reported in Table~\ref{tab:initial_data}.

The computational grids consist of a nested set of 10 refinement
levels, with the finest mesh having resolution $h=M/40$. This
resolution translates into a resolution of about $h=m/19 - m/12$, with
respect to the bare mass $m$ of the puncture according to
Table~\ref{tab:initial_data}. The minimal resolution found to be
adequate for spinning cases according to~\citet{Campanelli:2006uy} is
$h < M/30$.  In our $h=M/40$ simulations there are 4 refinement levels
of $58^3$ grid-points nested within 6 levels of $102^3$ grid-points.
During the evolution the shape and number of grid-points per
refinement level vary as the centers of the grids track the positions
of the black holes. The coarsest mesh is kept fixed and extends to
$650 M$ from the origin in each direction.

\section{Gravitational Recoil} 
\label{sec:recoil}

The gravitational recoil is computed from the rate of change of linear momentum
\begin{equation}
\frac{dP^i}{dt} = \lim_{r\rightarrow\infty}\left\lbrace
\int \frac{d^2E}{d\Omega dt} n^i \,r^2 d\Omega \right\rbrace \,,
\label{eq:dPdt}
\end{equation}
which is determined by the fluxes of energy $E$ and linear momentum $P^i$
($n^i$ is the unit normal to the sphere). In order to compute the
recoil velocity, the Newtonian momentum relation is used, $V^i=P^i/M$.

In terms of $\Psi_4$, the component of the Weyl curvature tensor representing
outgoing radiation, equation~(\ref{eq:dPdt}) reads \citep{1980grg2.conf....1N}
\begin{equation}
\frac{dP^i}{dt} = \lim_{r\rightarrow\infty}\left\lbrace
\frac{1}{4\pi}\int \left |\int^t_{-\infty}\Psi_4\,dt^\prime\right |^2 n^i \,r^2 d\Omega
\right\rbrace\,.
\label{eq:dPdtpsi4}
\end{equation}
Equation (\ref{eq:dPdtpsi4}) is applied at a finite radius $r>30M$
away from the ``center of mass'' of the binary but far enough from the
boundary of the computational domain to avoid the effects from
spurious reflection from the
boundary~\citep{Zlochower2005:fourth-order}.  The Weyl scalar $\Psi_4$
is computed in the bulk of the computational domain and is then
projected onto the sphere and used in the computation of
equation~(\ref{eq:dPdtpsi4}).

We also estimate the gravitational recoil using a mode decomposition.
Instead of constructing $\Psi_4$ in the bulk of the computational
domain and interpolating it on a sphere to be used in
equation~(\ref{eq:dPdtpsi4}), we decompose $\Psi_4$ into spin-weight
$-2$ spherical harmonics and then compute the recoil. That is, one
first constructs the coefficients $\,_{-2}C_{\ell m}$ such that
\begin{equation}
\Psi_4=\sum_{\ell m} \,_{-2}C_{\ell m}(t,r) \,_{-2}Y_{\ell m}(\theta,\varphi)\,.
\end{equation}
Given these coefficients, the gravitational recoil is given by
\begin{equation}
  \frac{dP^i}{dt}=\sum_{\ell m \bar\ell \bar m} 
\olap{\ell}{m}{\bar\ell}{\bar m}
\label{eq:recoil.modes}
\end{equation}
where $\olap{\ell}{m}{\bar\ell}{\bar m}$ represents the contribution to $dP^i/dt$
from the overlap 
\begin{equation}
  \olap{\ell}{m}{\bar\ell}{\bar m} \propto Re\left[ {}_{-2}\hat{C}_{\ell m}^\star \,_{-2}\hat{C}_{\bar \ell \bar m}
    \int n^i \,_{-2}Y_{\ell m}^\star {}_{-2}Y_{\bar \ell\bar m} d\Omega\right]\,,
\end{equation}
with $\,_{-2}\hat{C}_{\ell m}\equiv \int_{-\infty}^{t} \,_{-2}C_{\ell m} dt^\prime$.
This mode-overlap decomposition has the advantage that the 
contribution from different overlapping modes can be studied
individually. 

\begin{figure}
\plotone{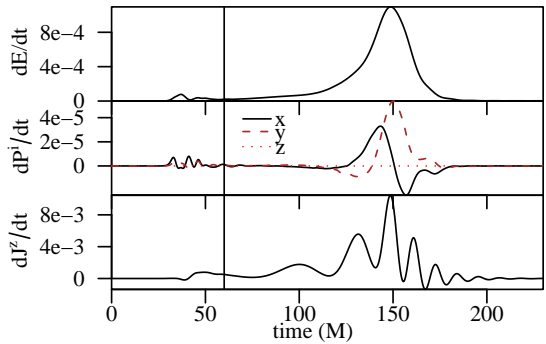}
\caption{\label{fig:fig1} Fluxes of energy $dE/dt$, linear momentum 
$dP^i/dt$ and angular momentum $dJ/dt$ as a function of time for the $S0.10$ 
($a=0.4$) case. The vertical
line at $60\,M$ denotes $t_{\mathrm{min}}$, the lower limit of the time integration
used to estimates kicks which avoids contamination from the spurious radiation in the
initial data.}
\end{figure}

There is an important issue to keep in mind when using both 
equations (\ref{eq:dPdtpsi4}) and (\ref{eq:recoil.modes}) to estimate
kicks. It is well known that initial data in \bbh{} simulations 
contain spurious radiation. Fortunately, this radiation does not
seem to have a significant effect on the dynamics of the binary. 
However, because of the time-integration involved in the kick formulas,
the estimates are affected by the spurious radiation.
To alleviate this problem, we set the lower limit in the
time integral to be $t_{\mathrm{min}}$ and choose
$t_{\mathrm{min}}$ as the time after which the spurious burst has passed.
As an example, Figure~\ref{fig:fig1} displays 
the fluxes of energy $dE/dt$, linear momentum $dP^i/dt$ and angular momentum $dJ/dt$ through the detector at
$r_{det}=30\,M$ for the $S0.10$ case. 
It is clear from these rates that there is a spurious burst from the initial data for $t<50\,M$.
In particular, notice the effect on  $dP^i/dt$ at early times.
The line at $t_{\mathrm{min}}=60\,M$ shows our choice for this cut-off. 
The precise choice of $t_{\mathrm{min}}$ is not important, as
long as the initial spurious burst is eliminated and $t_{\mathrm{min}}$
is not too close to the time when the amplitude of the
gravitational wave becomes relevant. Since we use several locations (``detectors'')
at different radii to compute fluxes, the value of $t_{\mathrm{min}}$
is adjusted as 
$t_{\mathrm{min}}=30\,M+r_{\mathrm{det}}$, where
$r_{\mathrm{det}}$ denotes the detector radius.
Note the smallness of  $dP^z/dt$ from Figure~\ref{fig:fig1}.
It translates to velocities of $\sim  0.2\,\KMS$; thus, we will not plot $V^z$ in subsequent figures.

\begin{figure}
\plotone{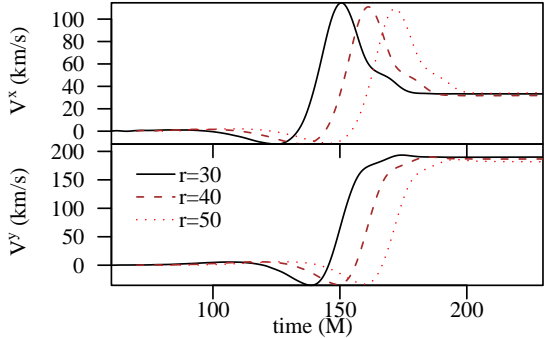}
\caption{\label{fig:recoil.detectors} Recoil velocity $V^x$ and $V^y$
  computed from different detector locations for $S0.10$ with resolution $h=M/40$.
  The detectors were located at $r_{\mathrm{det}}/M=(30,40,50)$.}
\end{figure}

Another important check when computing kicks using
equations (\ref{eq:dPdtpsi4}) and (\ref{eq:recoil.modes}) is 
the dependence of the results on the 
extraction radius $r_{\mathrm{det}}$. 
The kick formulas are in principle valid in the limit
$r \rightarrow \infty$, but one applies them at a finite
extraction radius $r_{\mathrm{det}}$ where there is sufficient resolution. 
Figure~\ref{fig:recoil.detectors} shows the recoil velocity as a function
of time computed at different detector radii,  $r_{\mathrm{det}}/M=(30,40,50)$.
The time dependence of the velocities has not been adjusted by 
the lag in arrival times at each detector. 
Although small, one can see from Figure~\ref{fig:recoil.detectors} that 
there a is slight sensitivity of the extracted kick velocity to the 
location of the detector for the ranges we considered.
This variation is within the error estimates of our kicks. 
The origin of this dependence of the extracted kick on the detector
location could be numerical (e.g.~outer boundary, mesh refinement interfaces, etc.)
or due to the redshift and tail effects.\footnote{We thank the anonymous referee
for bringing this to our attention.}
 
\section{Code Tests and Waveform Convergence}
\label{sec:converge}

We have tested that our code produces a sufficient level of
convergence for equal mass, non-spinning \bh{} binaries that we are
confident in the results. In particular, we have carried out extensive
tests~\citep{Shoemaker07} for the $R1$ run in \citet{Baker:2006yw}
and found resolution ranges that yield between 3rd- and 4th- order convergence.  
Also as a code test,
we carried out a non-spinning, unequal mass simulation for $\eta
=0.23$. The kick obtained from this run ($\sim 130 \KMS$) matches that
by \citet{Gonzalez:2006md}.  Because the \bbh{} setups in our present
work have no symmetries, the computational cost of each simulation is
high (for our $h=M/40$ resolution runs the cost is $\sim 44$ hours on $32$ CPU
cores for a total of about $\sim 1400$ CPU hours on a supercomputer),
so to demonstrate convergence our runs were limited to resolutions
$h\le M/40$. We present convergence results for the $S0.10$ case; the other
cases have similar behavior.

\begin{figure}
\plotone{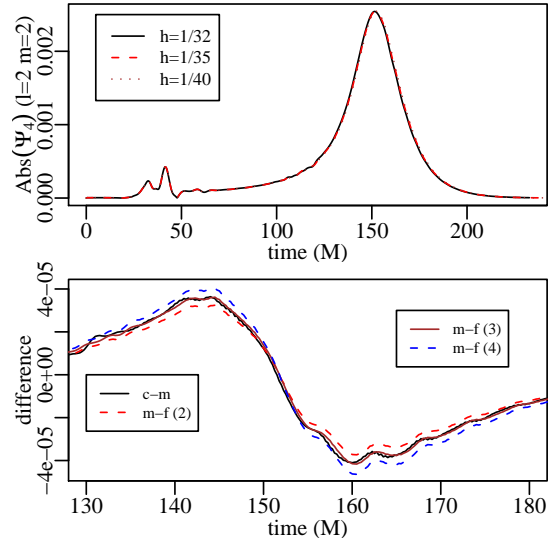}
\caption{\label{fig:fig3} The amplitude of the dominant $\ell=2,\, m=2$ mode of
  $\Psi_4$ for the case $S0.10$ ($a=0.4$). The top plot shows the mode at three different 
resolutions ($h/M =  1/32, 1/35, 1/40$), while
the bottom shows the small differences between the medium-coarse (``c-m'') and
the medium-fine (``m-f'') simulations rescaled for 2nd, 3rd and 4th order. The waveform 
is between 3rd- and 4th-order convergent.
}
\end{figure}

Figure~\ref{fig:fig3} shows the amplitude of the dominant $\ell=2,
m=2$ mode of $\Psi_4$. The top panel of the figure displays the mode
at the three different resolutions ($h/M = 1/32, 1/35, 1/40$). The
bottom panel shows the coarse-medium (``c-m'') differences and the
medium-fine (``m-f'') differences rescaled for 2nd, 3rd and 4th
order. As the plot shows, this mode converges between 3rd and 4th order. In our
convergence studies for other systems (e.g.~equal mass \bh{s}) getting
closer to 4th-order convergence required at least a factor of two
between the coarsest and finest resolution. Given the range of
resolutions that we are able to do for the present study, the
deterioration of our convergence should not be surprising.
Nonetheless, we believe that the observed level of convergence in our
simulations will not affect the astrophysical implications of the
magnitude of our kick estimates.

\begin{figure}
\plotone{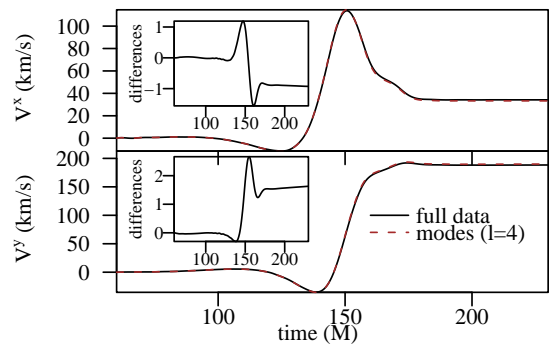}
\caption{\label{fig:recoil.mode.full} Recoil velocity $V^x$ and $V^y$
  versus time computed from equation~(\ref{eq:dPdtpsi4}) and
  equation~(\ref{eq:recoil.modes}) for the $S0.10$ model with resolution
$h=M/40$ extracted at $r_{det}=30\,M$. $V^z$ is below 0.2 km/s and hence is not
  shown. The insets labeled ``differences'' show the difference 
between the recoil from equation~(\ref{eq:dPdtpsi4}) and
  equation~(\ref{eq:recoil.modes}) with modes up to and including  $\ell=4$}.
\end{figure}

As a check of our implementation of the kick extraction,
Figure~\ref{fig:recoil.mode.full} compares the recoil velocity
computed from equation~(\ref{eq:dPdtpsi4}) and
equation~(\ref{eq:recoil.modes}) for the case $S0.10$ with 
resolution $h=M/40$. For equation~(\ref{eq:recoil.modes}), we include up to
$\ell=4$ modes. It is evident from this plot that with the modes  $\ell\le 4$
one can reconstruct most of the total recoil velocity.

\section{Results}
\label{sec:results}

First, we present the main results of our work, namely the kick estimates
together with the radiated energy and angular momentum, followed by a  discussion of 
convergence and a mode analysis of the kicks.

\subsection{Kicks and Radiated Energy and Momentum}

\begin{deluxetable}{l c c c c}
\tablewidth{0pt}
\tablecaption{\label{tab:radiated}Radiated Quantities}
\tablehead{
\colhead{Model} & \colhead{$a$} & \colhead{$V (\KMS)$} & \colhead{$\Delta E (\%)$} & \colhead{$\Delta J (\%)$} }
\startdata
S0.05 & 0.2 & $\phantom{0}96\pm 7$ & 3.24  & 26.82\\[1mm]
S0.10 & 0.4 & $190\pm 10$           & 3.30  & 27.05\\[1mm]
S0.15 & 0.6 & $285\pm 12$            & 3.33  & 27.12\\[1mm]
S0.20 & 0.8 & $392\pm 33$           & 3.34  & 26.83\\[1mm]
\enddata
\end{deluxetable}            

\begin{figure}
\plotone{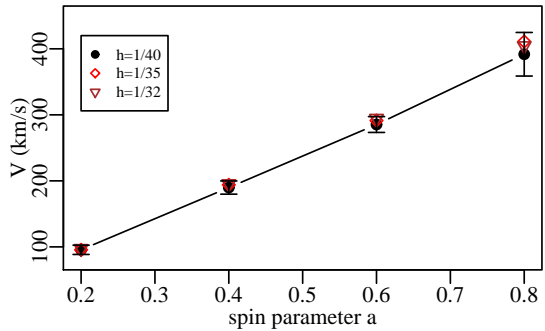}
\caption{\label{fig:recoil.spin.dep} Magnitude of the recoil velocity $V$
 as a function of the dimensionless spin parameter $a$. Solid circles are for resolutions
$h=M/40$, diamonds for resolutions $h=M/32$ and inverted triangles
for resolutions $h=M/30$. In each case, the results at different resolution 
cluster more tightly than the conservatively estimated error 
bars (Table~\ref{tab:radiated}). }
\end{figure}

The core results of our work are summarized in Table~\ref{tab:radiated}.
Table~\ref{tab:radiated} lists the values for the total
recoil $V$, energy $\Delta E$ and angular momentum $\Delta J$ radiated 
for each of the cases considered. The reported values
were obtained with resolutions $h=M/40$ and extracted at $r_{det} = 40\,M$. 
For reference, we include also the dimensionless spin parameter $a$.
Figure~\ref{fig:recoil.spin.dep} displays the recoil
velocity $V$ as a function of the dimensionless spin parameter $a$ for all the resolutions 
used in our simulations.
Solid circles denote resolutions
$h=M/40$, diamonds resolutions $h=M/32$ and inverted triangles
resolutions $h=M/30$.
The error bars correspond to the conservatively estimated errors 
listed in Table~\ref{tab:radiated}, and are 
larger than the actual scatter of the results at different resolution.

In order to estimate these errors, for each spin case, 
we perform Richardson error estimates of the total
recoil velocity $V$ assuming 2nd order convergence.
We then increase these errors to
take into account factors such as
the deterioration of convergence in the weak mode-overlaps 
(see below).
We believe these are conservative best-guess errors 
that could be reduced with, among other things, higher resolution.

Note in Figure~\ref{fig:recoil.spin.dep}
the linear dependence of 
the magnitude of the kick velocity $V$ on the spin parameter,
as expected from the multipole example in Section~\ref{sec:intro}.
A fit to the data yields \Vkick.

An interesting aspect of the spin configuration we have considered is
the fraction of radiated energy $\Delta E$ and angular momentum $\Delta
J$. The fraction radiated is approximately constant within the accuracy 
of our simulations.  One possible reason why $\Delta E$ and $\Delta J$
do not seem to depend on the spins of the holes could be due to the
set up of our initial data.  By construction, the four
cases we considered have the same total initial angular momentum
$J/\mu\,M = 3.3$. In our case with spins oppositely directed and
with equal magnitude the variations in the total ADM energy are $<
0.05\%$, as can be seen from Table~\ref{tab:initial_data}.

\subsection{Mode Analysis and Convergence}

\begin{figure}
\plotone{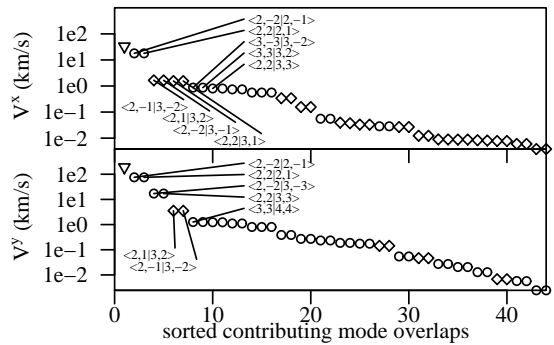}
\caption{\label{fig:recoil.mode.sorted} 
Contribution to the recoil velocity components $V^x$ and $V^y$
from each $\olap{\ell}{m}{\bar\ell}{\bar m}$ mode-overlap 
for the $S0.10$ case with resolution $h=M/40$ extracted 
at $r_{det}=30\,M$. 
The recoil from combining all mode-overlaps is labeled with an inverted triangle.
Positive mode-overlap contributions are labeled with circles and negative with diamonds.}
\end{figure}

With the kick formula (\ref{eq:recoil.modes}), we were able to
investigate the contribution of each mode-overlap
$\olap{\ell}{m}{\bar\ell}{\bar m}$ to the total recoil velocity.
Figure~\ref{fig:recoil.mode.sorted} shows the contribution that each
mode-overlap makes to the total kick velocity for the $S0.10$ case
with $h=M/40$ resolution.  The mode-overlaps have been sorted from
largest to smallest.  The total recoil is labeled with an inverted
triangle.  Positive mode-overlap contributions are labeled with
circles and negative with diamonds.  There are two important points to
take from this figure:  A) Note how quickly the contribution to $V^x$ and
$V^y$ from each mode-overlap falls off; that is, there are few mode-overlaps
that have significant contribution. B) The two
most dominant mode-overlaps $\olap{2}{-2}{2}{-1}$ and
$\olap{2}{2}{2}{1}$ contribute almost equally $54\,\%$ (note that
other modes contribute negatively) in $V^x$ and $40\,\%$ in $V^y$.

\begin{figure}
\plotone{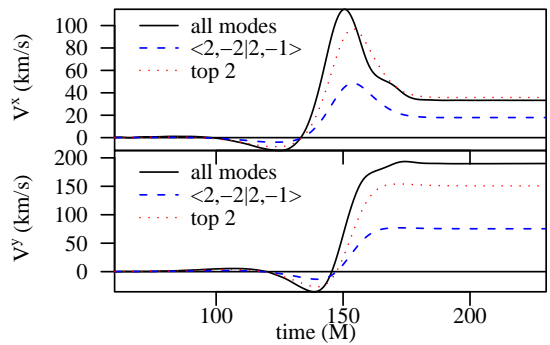}
\caption{\label{fig:recoil.mode.acc} Recoil velocity components $V^x$ and $V^y$
versus time for the case $S0.10$ case with resolution  $h=M/40$ extracted at $r_{det}=30\,M$.
The solid line gives the accumulation in time of recoil from all mode-overlaps combined,
dotted line denotes the combined accumulations of only the two most dominant mode-overlaps,
$\olap{2}{-2}{2}{-1}$  and $\olap{2}{2}{2}{1}$, and the dashed line
the accumulation in time of the $\olap{2}{-2}{2}{-1}$ mode-overlap.}
\end{figure}

Another way of showing the dominance of the
$\olap{2}{-2}{2}{-1}$  and $\olap{2}{2}{2}{1}$
mode-overlaps is presented in Figure~\ref{fig:recoil.mode.acc}.
This figure shows the accumulated velocity as a function of time.
The solid line gives the accumulation in time of recoil from all mode-overlaps combined, the
dotted line shows the combined accumulations of the two most dominant mode-overlaps,
$\olap{2}{-2}{2}{-1}$  and $\olap{2}{2}{2}{1}$, and the dashed line displays
the accumulation in time of the $\olap{2}{-2}{2}{-1}$ mode overlap.

\begin{figure}
\plotone{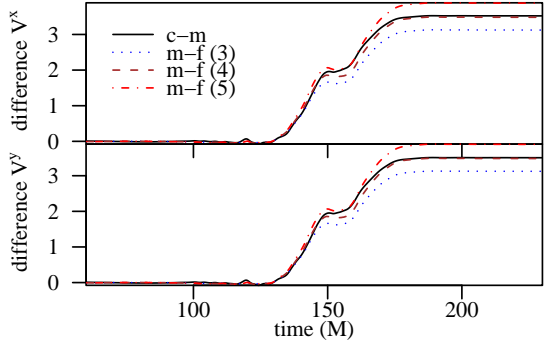}
\caption{\label{fig:recoil.mode.conv} Convergence analysis of the recoil
  contribution from the dominant overlap $\olap{2}{-2}{2}{-1}$ for the
$S0.10$ case extracted at $r_{det}=30\,M$.
The solid line gives the difference between the coarse and medium resolutions (``c-m'').
The other lines show the difference between the medium and fine resolutions (``m-f''),
scaled to match (``c-m'') for 3rd, 4th and 5th order convergence.
}
\end{figure}

Given that the mode-overlaps $\olap{2}{-2}{2}{-1}$  and $\olap{2}{2}{2}{1}$
are the principal contributors to the total kick velocity,
we analyzed the convergence properties of these overlaps.
Figure~\ref{fig:recoil.mode.conv} displays the differences of the 
$\olap{2}{-2}{2}{-1}$ mode-overlap from three resolutions,
 $h/M=( 1/32, 1/35, 1/40)$. The 
solid line is the difference between the coarse and medium resolutions (``c-m'').
The other lines show the difference between the medium and fine resolutions (``m-f''),
scaled to match (``c-m'') for 3rd, 4th and 5th order convergence.
It is clear from this figure that this mode-overlap is close to being
4th-order convergent. A similar situation occurs for the other
dominant mode-overlap $\olap{2}{2}{2}{1}$.
Unfortunately, the situation is different for the other weaker
mode-overlaps. These overlaps involve 
higher modes of $\Psi_4$ that are much more difficult to resolve given 
the range of resolutions we have. 
When these weaker modes are added to obtain the total recoil,
one is no longer able to reach the desired 4th-order convergence. 
In some instances it drops to 1st-order convergence.
Fortunately, as we have seen from Figure~\ref{fig:recoil.mode.sorted},
their contribution to the overall recoil is small.
We are confident that our
total kick velocities will not change significantly if one is able to
achieve finer resolutions than $h=M/40$.

\section{Summary and Discussion}\label{sec:discussion}

We have computed estimates of \bh{} merger kick velocities from 
previously untreated physical effects arising from the spin of the holes. 
Our computational simulations provided firm predictions of kick velocities
for \bbh{} systems of equal mass and anti-aligned spins. 
Because we are able to accurately resolve the dominant modes 
that contribute to the kick and estimate those kicks by a number of methods, 
we are confident in our astrophysical conclusions involving the binary types we considered.
Previous studies which considered the merger of (non-spinning) \bh{s} of unequal masses 
produced kicks $\sim 200\,\KMS$ with a reasonably broad maximum 
near the symmetrized mass ratio of $\eta=0.2$ (mass ratio $0.38$). 
From the astrophysical point of view, $200\, \KMS$ is interesting. For instance,
the escape velocity from the center of dwarf elliptical galaxies is $~300~\KMS $, 
assuming the standard picture of dark matter halos.
We found spin kick velocities \Vkick, where $a$ is the dimensionless spin parameter, 
in opposite-spin configurations (see Figure~\ref{fig:recoil.spin.dep}). 

For black holes ($10 - 20\, M_\odot$) seen in the galaxy,
there are observations supporting spin parameters $a \gtsim 0.8$~\citep{2006ApJ...652..518M}, 
and theoretical explanations of why this is so 
are generally applicable to \smbh{s} also.  
Thus we expect substantial kicks due to spin interactions. 
Our simulations predict typical kicks $\gtsim 400 \,\KMS$ in astrophysical \bh{} mergers of all masses.
These results could explain the observed 
absence of central black holes in dwarf elliptical galaxies.
Our simulations show limitations, mostly due to the high cost of
performing very high resolution runs. But, because we are 
able to accurately resolve the dominant modes that contribute to the kick,
we believe that our astrophysical conclusions are secure\footnote{Soon after the completion of our work, 
results that support our findings of spin effects on kicks were obtain by~\citet{2007gr.qc.....1164C, Koppitz:2007ev}.}.

\acknowledgments
Thanks to Jos\'e Gonz\'alez, Ben Owen, Carlos Sopuerta, Ulrich Sperhake and Nico Yunes 
for helpful conversations.
The authors acknowledge the support of the Center for
Gravitational Wave Physics funded by the National Science Foundation
under Cooperative Agreement PHY-0114375. This work was supported by
NSF grants PHY-0354821 to Deirdre Shoemaker, PHY-0244788 and
PHY-0555436 to Pablo Laguna and PHY-0354842 and NASA grant NNG 04GL37G
to Richard Matzner.  Computations were carried out at NCSA under
allocation TG-PHY-060013N, and at the Texas Advanced Computation Center, University of Texas System.

\bibliography{references}

\end{document}